\documentclass[runningheads]{llncs}
\usepackage[T1]{fontenc}

\usepackage{graphicx}

\usepackage{orcidlink}
\usepackage{todonotes}
\usepackage{microtype}

\begin{document}

\title{A Small Dataset May Go a Long Way:\\ 
Process Duration Prediction in Clinical Settings%
}

\titlerunning{Improving Duration Prediction of Clinical Processes}
\author{Harald Störrle\inst{1}\orcidlink{0000-0003-2755-599X} \and
Anastasia Hort\inst{1}\orcidlink{0009-0001-6053-9804} }
\authorrunning{H.\ Störrle, A.\ Hort}

\institute{QAware GmbH, Munich, Germany,\\
\email{\{harald.stoerrle, anastasia.hort\}@qaware.de}\\
\url{https://www.qaware.de/}}
\maketitle           
\begin{abstract}

\textbf{Context:} Utilization of operating theaters is a major cost driver in hospitals. Optimizing this variable through optimized surgery schedules 
may significantly lower cost and simultaneously improve medical outcomes. Previous studies proposed various complex models to predict the duration of procedures, the key ingredient to optimal schedules. They did so perusing large amounts of data.

\textbf{Goals:} We aspire to create an effective and efficient model to predict operation durations based on only a small amount of data. Ideally, our model is also simpler in structure, and thus easier to use.

\textbf{Methods:} We immerse ourselves in the application domain to leverage practitioners expertise. This way, we make the best use of our limited supply of clinical data, and may conduct our data analysis in a theory-guided way. We do a combined factor analysis and develop regression models to predict the duration of the perioperative process.

\textbf{Findings:} We found simple methods of central tendency to perform on a par with much more complex methods proposed in the literature. In fact, they sometimes outperform them. We conclude that combining expert knowledge with data analysis may improve both data quality and model performance, allowing for more accurate forecasts.

\textbf{Conclusion:} We yield better results than previous researchers by integrating conventional data science methods with qualitative studies of clinical settings and process structure. Thus, we are able to leverage even small datasets.

\end{abstract}

\section{Introduction}

Conducting surgical procedures is one of the core processes of hospitals with a paramount impact on the economic balance sheet, accounting for an average of 40\% of all expenses and generating approximately 60\% of total revenue. The perioperative process stretches from the moment a procedure is decided upon to the moment where the patient is independent of medical supervision after a surgery again \cite{verywellhealth2025}. 

\begin{figure}[tb]
\begin{center}
    \includegraphics[width=\textwidth]{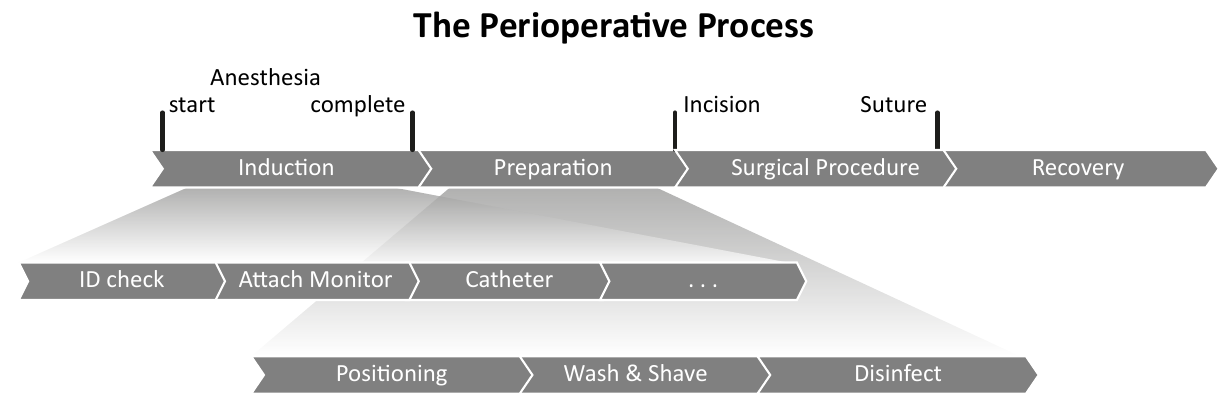}
    \caption{The perioperative process is divided in the preoperative steps, commonly subsumed as induction, a surgical procedure, and recovery. Induction comprises the administration of anesthesia, positioning of the patient to facilitate access to body regions relevant to the surgical procedure, and further preparatory steps. From a process management point of view, the single largest independent variable is the precision of the duration estimation. Surgery duration is commonly measured as the incision-suture-time.}
    \label{periop_processes}
\end{center}
\end{figure}

The perioperative processes \cite{hughes2022oxford} are concerned with activities before, during, and after a surgical procedure. They are critical from an economic point of view, because they take place in operating rooms, one of the most resource-intensive and strategically important areas of clinical infrastructure. The perioperative process may differ between hospitals, reflecting variations in management structure, clinical routines, and technical systems. In this study, we focus on three main sub-processes of the perioperative workflow at the LMU University Hospital, see Fig.~\ref{periop_processes}. Because of their central role, we now describe them in greater detail.

During the \emph{Induction}, the patient is being moved to the operating room, their identity is verified, monitoring devices for heart rate and respiration are attached, and catheters are emplaced. Most importantly, anesthesia is being administered. The duration of the induction phase is the time between start and completion of anesthesia. %
After induction, \emph{Preparation} starts, where the patient is positioned according procedural requirements, and the incision area is being prepared by washing, shaving, and disinfecting the skin, as appropriate. Often, a pause ensues while waiting for the surgeons. %
Then, the \emph{Surgical Procedure} proper may start, marked by the \emph{Incision} time. Surgery ends with the final suture. The duration of the surgical procedure is defined as the time between incision and the last suture. In the recovery phase, the anesthesiologist remains with the patient until they begin to breathe independently. 

Surgical departments create daily or weekly operation schedules determining exactly when which operation is to take place, which medical staff and other resources need to be available at what time, and when patients are admitted to the wards for intake and examination. Creating optimal operation schedules is instrumental for hospitals: If schedules are too loose, the operating rooms are underutilized and create an economic burden. On the other hand, if they are too tight, operations have to be deferred, which incurs costs that are unrecoverable from health insurers. Also, deferring operations may increase patient anxiety, while extending shifts may decrease staff satisfaction. So, optimizing operation schedules to improve the perioperative process offers promises significant reductions in healthcare costs, improved quality of medical services, and working conditions for medical staff \cite{ucocorcmg2018,orerr2018,orpscszf2019}.

However, surgical planning is an inherently complex process involving specialists from various departments, from surgeons and anesthesiologists across multiple specialties, via nursing staff to operating room coordinators and hospital administrators. All of these stakeholders must work in harmony to ensure a successful outcome. Additionally, emergency surgeries and unexpected patient reactions to medication may occur at any time, and take absolute precedence over any preconceived plan. Quick adaptation and restructuring of the surgery schedule occurs frequently at the LMU (though it might be less frequent in other hospitals). All in all, the planning process is highly dynamic and constantly evolving, requiring flexibility and high degree of coordination \cite{osssdudvv2007,spsedc2011}. From a process management point of view, this leaves us with two avenues of improvement. First, we may create optimal plans that, in the absence of any perturbations from emergencies, allow accurate and stable plans. Second, by creating such plans automatically, allowing for ad-hoc replanning, we may be able to mitigate or minimize the deleterious effects of emergencies. In this study, we show how both of these may be achieved by analyzing clinical processes and extracting high-quality predictive models from them. 

\section{Related Work}
\label{related_work}

Recent studies have focused on using machine learning techniques to better predict how long surgeries will take. The goal is to schedule operating rooms more efficiently, reduce patient wait times, and make the best use of hospital resources. Even though the studies used datasets from different clinics, focused on various factors, and applied a range of modeling methods, they still reveal recurring patterns and consistent findings.

The analysis of existing research reveals that a critical step in all studies is the careful data preprocessing, which directly influences the accuracy of predictions. Kendale et al. \cite{mlppcddulkbb2023} excluded stop words and applied TF-IDF to standardize procedure names, while Martinez et al. \cite{mlstpmmpr2021} used One-Hot and ordinal encoding. Yuniartha et al. \cite{esmpsduymh2021} enhanced features with comorbidity and allergy information. The range of methods used in modeling varied from regression techniques to ensemble methods, with Random Forest and Gradient Boosting Machine (especially XGBoost) repeatedly showing superiority \cite{elaipcgms2023,mlppcddulkbb2023,umlporcsmgc2023,dpmscdupr2025,ippsdrhk2023}, and Bagged Trees demonstrating high accuracy and efficiency \cite{mlstpmmpr2021}. Neural networks, though less commonly applied, as in the work of Jiao et al. \cite{crtpsmaj2022} with MANN based on LSTM, were able to account for temporal structure and outperformed Bayesian models in real-time prediction. Adaptive solutions were also developed: the hybrid model by Soh et al. \cite{soh2020hybrid} used different regression algorithms for specific data subsets, taking into account the dynamic nature of the clinical environment, while Hosseini et al. \cite{hosseini2015surgical} demonstrated the effectiveness of stepwise regression for certain surgical specialties, highlighting procedure codes as the main factor in prediction.
Evaluation metrics were largely consistent in most studies, with mean absolute error (MAE) and root mean square error (RMSE) serving as standard benchmarks. Additional metrics such as $r^2$, MAPE, and median absolute deviation were employed to capture different aspects of model accuracy and robustness.

Finally, the integration of domain knowledge was a distinguishing factor in the study by Strömblad et al. \cite{effpmps21}, who collaborated with clinical staff to identify over 300 relevant predictors. Although implementation details were not thoroughly reported, the inclusion of expert insights significantly contributed to improved scheduling outcomes.

Numerous studies aimed at improving the accuracy of surgical planning rely on the analysis of large datasets (typically 3-4 years' worth of data). In contrast, our dataset represents only one year of work (see Section~\ref{data_prepare} for details). The small sample size made it difficult to model data for individual departments. In response, we propose a new approach that involves modeling not only surgeries but also the entire perioperative process. By breaking the process into several stages and integrating knowledge about its specifics, we identified key factors affecting the duration of each step. This approach makes planning more flexible and adaptive to unforeseen situations.

\section{Research Methods}
\label{methods}

Our approach combines quantitative and qualitative methods. First, we collected viewpoints of the experts who engage with these processes every day. Second, we processed data from a clinical information system to validate, test, and quantify the expert opinions gathered before.

In the qualitative part of our study, we conducted semi-structured interviews with selected medical staff applying theoretical sampling \cite{breckenridge:jones:2009:demystifying,denzin:lincoln:2018:handbook,glaser:strauss:1969:discovery,saldana:2011:fundamentals}. In this phase, we sought  to understand which factors they believed influenced the duration of various stages. These qualitative findings resulted in a causal model from which we derived hypotheses and requirements for the data we acquired.

In the quantitative part of our work, we processed these data and applied various statistical methods to test our hypotheses. Depending on the data properties, we applied the t-test, F-test, and the Kruskal--Wallis test to identify significant differences between groups and to explore potential causal relationships\footnote{Background information on the applied statistical tests can be found in \cite{MurphyMyors2023}.}. 

We also used the data to generate predictive models for procedure duration. Compared to similar studies, our dataset is relatively small in terms of both the number of documented surgeries and the number of features describing them. This posed difficulties for unsupervised learning due to sparsity and limited feature diversity.

This hybrid approach enabled us to understand the surgical workflow not merely as a sequence of chronological events, but as a complex interaction of medical, logistical, and human factors. It led to a deeper understanding of how processes actually unfold, where uncertainties arise, and how these can be actively managed. Ultimately, the model served as the foundation for data-driven predictions that are not only accurate but also interpretable and trustworthy, an essential step toward adaptive, transparent, and user-centered surgical process planning.

\section{Qualitative Analysis}
\label{qualitative}

We collected viewpoints of the experts who engage with these processes every day, including a diverse set of medical practitioners, including anesthesiologists, surgeons, and nurses. We conducted individual semi-structured interviews with a follow-up survey, and derived a causal model about influence factors, likely causal relationships and potential outcomes. Our insights are visualized in Fig.~\ref{factor_model} below. 

\begin{figure}[tb]
\begin{center}
    \includegraphics[width=\textwidth]{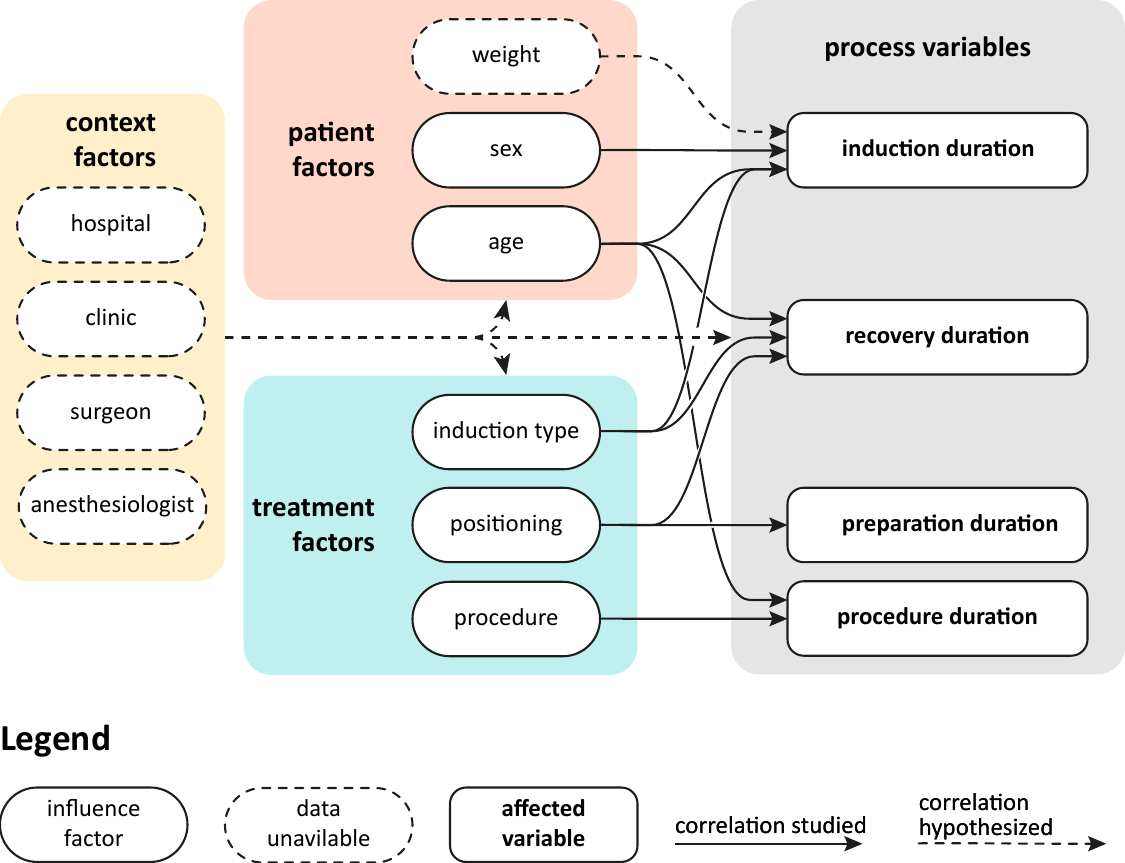}
    \caption{Potential influencing factors: This figure illustrates the factors and their possible impact on the various steps of the perioperative process. All mentioned factors and their influence are based on insights gained through communication with clinical experts.}
    \label{factor_model}
\end{center}
\end{figure}

These hypotheses formed the foundation of a causal model that not only reflects data-based dependencies but also structures and reveals the implicit knowledge embedded in clinical practice. The model was developed iteratively and validated in collaboration with clinical staff, with the goal of creating a realistic and practical representation of everyday operations--clear, transparent, and aligned with clinical routines.

We identified four relevant process variables: the durations of induction, preparation, the surgical procedure proper, and recovery. In our study, we focus on induction and procedure duration, which are measured as follows:

\begin{itemize}
    \item Induction Duration is the difference between the timestamps  \texttt{anesthesia\_start} and \texttt{anesthesia\_complete}.
    \item Surgery Duration is the difference between the timestamps \texttt{incision} and \texttt{suture}.
\end{itemize}

These four variables may be under the influence of three groups of factors. First, there are patient factors such as age, sex, and weight. Clearly, patient age affects many physiological and epidemiological variables, so that age affects almost all variables of the perioperative process. For instance, in old age, recovery from anesthesia takes longer. Also, there are many procedures that affect different age groups in a different way. For instance, hip replacements are very rare in younger patients. 

Second, there are treatment factors, that is, factors pertaining to the medical practice. For instance, obviously, the surgical procedure performed has a major influence on the duration of the surgery. But this influence extends to the preoperative phase, for instance in that the positioning and induction depend on it as well. Observe, that there are cross-dependencies, too, such as the correlations between sex and procedures (e.\ g., mastectomy affects women almost exclusively). 

Third, there are context factors such as the hospital, the clinic within the hospital, and the individual surgeons in a given clinic. Our informants emphasized the influence of individual capabilities of a surgeon on surgery quality and duration. Also, the kind of procedures performed varies greatly with clinic. E.\ g., only a neurology clinic will perform brain surgery, which is much less standardized than, say, many orthopedic procedures.

We did not study second-order dependencies between patient and treatment factors, as they present redundant information only. Also, we did not study patient weight, as this information was not contained in our data set. However, this would make for interesting follow-up work, as we suspect that patient weight has a major influence on induction, e.\ g., by affecting the duration of positioning. Similarly, we did not study the influence of context factors, as our data only affects one hospital, and data an clinics and individual surgeons was not available to us. However, again, this would make for interesting follow-up work as it may uncover medical best practices that could be transferred from one hospital to another.

\section{Quantitative Analysis}
\label{quantitative}

Besides the qualitative study, we modeled the perioperative process based on the collected data. Using classical process mining tools from the \textit{PM4PY} library, we reconstructed medical workflows from log data. The recovered workflows were divided into sub-processes, and for each one, we examined the relationship between predictors, such as patient positioning or anesthesia type, and process duration. This provided a structured basis for creating predictive models.

\subsection{Data Preparation}
\label{data_prepare}

We received data about operation planning and actual operations from a collaboration partner, Sqior Medical GmbH (henceforth Sqior, see \url{www.sqior.de}), a leading producer of health care information systems, who equips the LMU University Hospital Großhadern (henceforth LMU) one of the largest university hospitals in Germany. The data covered all operations taking place at the LMU between January 18, 2024 and January 23, 2025, and includes detailed documentation across a range of specialized departments: otolaryngology, gynecology, cardiac surgery, urology, neurosurgery, orthopedics, plastic surgery, visceral surgery, thoracic surgery, and vascular surgery. This broad coverage of specialties allows for an analysis of processes and the identification of planning and operational characteristics across different areas of medicine. The dataset consists of 427,959 events across 23,687 workflows, of which 17,358 include both incision and suture times. For the purpose of this study, only workflows containing both incision and suture timestamps were considered.

\begin{figure}[tb]
\begin{center}
    \includegraphics[width=.6\textwidth]{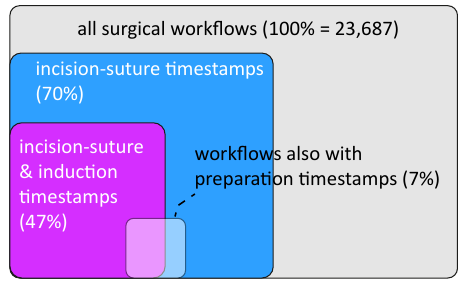}
    \caption{Our base dataset contains all documented workflows recorded in the Sqior system from January 18, 2024 till January 23, 2025. Many records were incomplete and did not contain all timestamps necessary. Therefore, we created sub-sets for different analyses.}
    \label{data_overview}
\end{center}
\end{figure}

The data we received contained up to 26 distinct events for each procedure. In collaboration with the clinic and Sqior we focused on three key sub-processes: \textit{induction}, \textit{preparation} and \textit{procedure}. The duration of these steps was calculated using the time difference between the following events: \textit{anesthesia\_start}, \textit{anesthesia\_complete}, \textit{incision}, and \textit{suture}. Additional fields include the planned duration for both induction and procedure, the department (such as orthopedics or neurology), and patient-related information such as age and sex. The data for each sub-process was aggregated into separate datasets. Each set contained the duration of the analyzed sub-process within each workflow, along with the relevant factors. Fig.~\ref{data_overview} illustrates the total number of workflows obtained, as well as the distribution across each sub-process.

After calculating the duration for each sub-process, records with missing or implausible values were removed. Subsequently, we relied on expert knowledge from clinical staff. We applied the $1.5\times$ interquantile range (IQR) \cite{LeveneHarris2025} method. Process durations falling outside this range were excluded from further analysis.

As a result, the dataset included 11,296 unique procedure descriptions from 16,576 records, as well as 2,081 unique induction (anesthesia) descriptions from 11,076 records. Firstly, we removed all non-alphanumeric characters from the text and replaced them with an empty string. The resulting text data was converted to lower case. To prepare the induction data, we applied normalization techniques supported by medical knowledge and resources from Sqior, including the unification of synonyms and abbreviations. 

The text data was transformed into numerical representations using TF-IDF vectorization. To cluster descriptions, we applied K-Means for procedures and Gaussian Mixture Models (GMM) for induction data. The number of clusters was determined with the mean Silhouette Coefficient.

The dataset was split into training (80\%) and test (20\%) sets. One-Hot Encoding was applied for categorical variables such as sex, department, and Target-Encoding (with smoothing parameter 40) for procedure clusters. To reduce the risk of overfitting, we implemented Smoothed Target Encoding based on Micci-Barreca's method \cite{10.1145/507533.507538}. The target value was computed based only on the training set and then assigned to the corresponding clusters in the test set to ensure that no information from the test set influenced the encoding process.

\subsection{Data Analysis}
\label{data_analysis} 

To analyze textual predictors such as induction type and anesthesia method, we used TF-IDF \cite{Papineni2001} vectorization. This enabled us to apply unsupervised learning techniques such as K-Means \cite{HartiganWong1979} and GMM \cite{10.1111/1468-0262.00273,10.1145/3408318} to form clusters of semantically similar descriptions. A hybrid approach that combined algorithmic segmentation with clinical expertise allowed us to group interventions by complexity and structure in a clinically meaningful way.

We then evaluated predictive performance using both simple methods, such as calculating the mean, and more complex regression models including linear regression \cite{James2023}, Random Forest \cite{Breiman2001}, and Gradient Boosting Machines (GBM) \cite{he2019gradientboostingmachinesurvey}. All models were tuned using a grid search.

Model evaluation was performed using different combinations of predictors, such as anesthesia type, patient positioning, age, sex, and cluster membership. To assess prediction accuracy, we used metrics such as MAE and mean percentage deviation from the plan. Special attention was given to model robustness under data variability and its practical applicability in clinical planning. Importantly, a deviation of ±20\% from the scheduled duration is considered acceptable in the clinical setting, and this threshold was incorporated into model assessment.

\section{Observations and Findings}
\label{observ_find}
At the beginning of our analysis, we focused on the most relevant metric in the dataset: the incision-to-suture time, which serves as a key indicator of actual surgical duration. Descriptive statistics revealed systematic over- and underestimation in manual planning. Surgeries planned for less than fifteen minutes were almost always underestimated, as shown by box-plots where the entire deviation distribution fell below zero. Overall, more than 60\% of surgeries deviated by over 20\% from their planned duration, with an average deviation of 68\%. These results clearly show that current planning is often inaccurate and that there is a strong need for data-driven improvements.

To standardize and reduce variability in free-text fields efficiently, we simplified the language by reducing complex expressions and applying stemming. Using TF-IDF vectorization and K-Means clustering, we grouped similar descriptions into meaningful categories. The optimal number of clusters was identified using the silhouette method. Anesthesia descriptions contained significantly more abbreviations than surgical ones. Here, Sqior’s internal algorithms helped standardize terms, reducing the number of clusters from 135 to 15 without any loss in prediction quality. These 15 clusters are more interpretable and manageable in clinical practice.

We conducted a statistical analysis using t-tests, ANOVA, and Kruskal--Wallis tests to evaluate the influence of different factors on duration. While demographic variables such as age and gender were statistically significant, their practical effect was minimal. In contrast, the type of procedure, anesthesia method, and patient positioning were both statistically and practically relevant. Exploratory factor analysis confirmed that the most influential variables for surgical duration were the procedure description and department, while for anesthesia, the description of the method played the key role. Gender had no significant impact.

\begin{figure}[tb]
\begin{center}
    \includegraphics[scale=0.48]{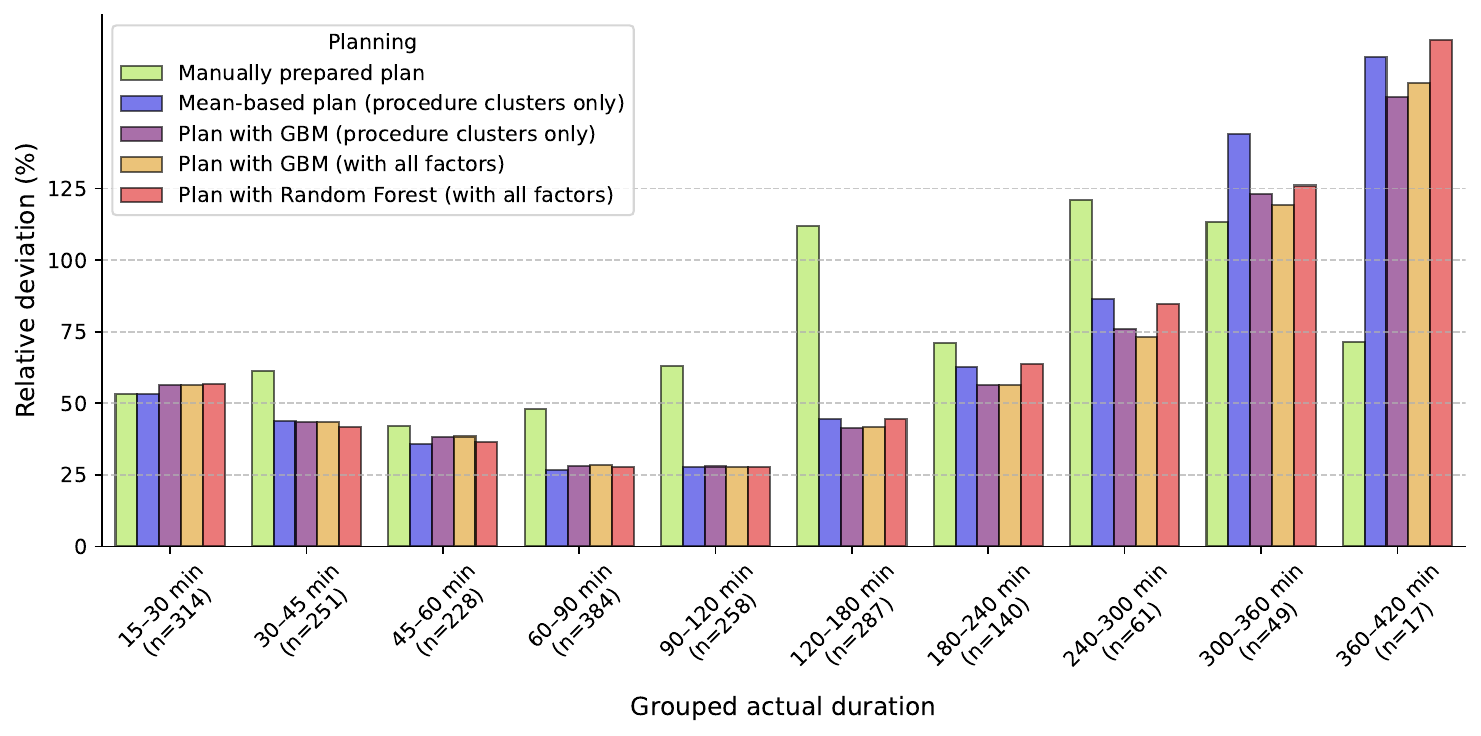}
    \caption{Comparison of different planning approaches for predicting the duration of surgical procedures. Four strategies were evaluated: manual planning, calculation based on the arithmetic mean, Random Forest, and GBM. The relative percentage deviations were analyzed in relation to the actual durations of the surgeries. Both the arithmetic mean and GBM approaches used the surgical procedure description as input. In addition, patient age was included in the GBM and Random Forest models. However, the predictions generated by the GBM model showed no statistically significant differences of whether age was included. This indicates that in this context, patient age has little influence on the duration of the surgery.}
    \label{ml_performance}
\end{center}
\end{figure}

These findings were incorporated into regression models, ranging from simple averages to more advanced approaches such as linear regression, Random Forest, and GBM. We found that prediction accuracy remained relatively stable across different combinations of features and models. This aligned with the statistical findings, which showed that complex models offered only marginal improvements, while the simple mean already provided robust predictions (see Fig.~\ref{ml_performance}). Given the high effort required to tune hyperparameters in complex models, the mean was chosen as a practical baseline solution.

Unlike traditional methods that estimate procedure duration based on extracted procedure name or surgeon's subjective assessments, our approach leverages semantic clustering of more than 11,000 unique free-text procedure descriptions. We grouped procedures based on semantic similarity rather than literal string matches. This allowed us to calculate stable average durations within clusters. This strategy is particularly valuable because, in real clinic practice, the same procedure can be described in many different ways depending on the surgeons' style or preferences. The traditional method of averaging by exact procedure name, like the Method of Taking Averages (MTA), does not account for this linguistic variability, resulting in inconsistent statistics and reduced reliability of estimates. In contrast, our approach creates a generalized and more robust model that handles lexical variation without sacrificing accuracy \cite{mlppcddulkbb2023,soh2020hybrid,effpmps21,esmpsduymh2021}. Moreover, this clustering technique provides the foundation for developing a new catalog of surgical procedures that better reflects the real-world diversity of medical documentation within the hospital. Such a catalog could significantly improve standardization, planning accuracy, and interoperability across departments and systems.

Furthermore, our study demonstrates that using a relatively small but carefully selected set of predictors such as induction type, patient positioning, and procedure type, together with simple models, led to a significant improvement in prediction accuracy compared to the original planned values. The reduction in average error was especially noticeable, with a decrease of 7.94 percentage points for the induction phase and 18.32 percentage points for the surgical phase.

Our analysis also carries practical implications. Automatic minimum duration recommendations, such as at least twenty minutes for induction, and buffer recommendations for high-variance clusters, can help improve planning accuracy. Early warning systems could identify bottlenecks when too many long procedures are scheduled in the same time window. 

Nonetheless, there are limitations. The dataset lacks key contextual variables, such as the availability of personnel, nursing staff levels, or patient-specific risk factors like comorbidities or catheter-based imaging workflows. These gaps limit the explanatory power of the models. In addition, small cluster sizes (less than fifty cases) in sensitive areas reduced statistical reliability. Future work should integrate additional data sources and apply expert-based cluster validation to further enhance the precision and robustness of the models.

\section{Interpretation}
\label{interpretation}

\begin{figure}[tb]
\begin{center}
    \includegraphics[scale=0.49]{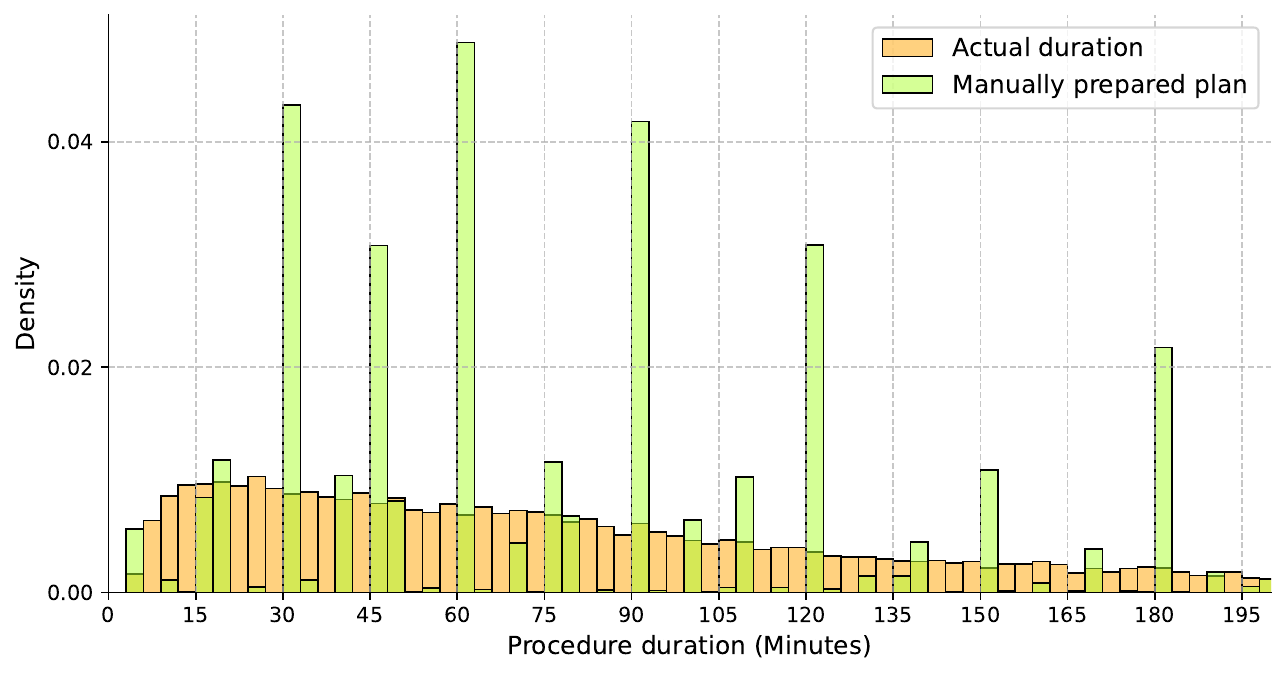}
    \caption{Distribution of actual surgery duration and the manual plan (3-minute interval)}
    \label{actual_state}
\end{center}
\end{figure}

We observed that manually scheduled operations are typically estimated at durations that are multiples of 15 minutes (see Fig.~\ref{actual_state}). The true operation durations, however, are smoothly distributed over time. We strongly suspect that medical staff estimate operation durations in an unsystematic way, based on intuition rather than data. Very likely, not much effort or attention is given to duration estimations. Possibly, they are seen as ephemeral and unimportant. 

We have shown that much more precise duration estimates can be generated automatically, that is, automation yields higher accuracy at lower planning cost. Clearly, more precise duration estimates are conducive to increased plan stability, and thus offer significant improvement of hospital economics and medical outcomes.

Previous studies have focused exclusively on the procedure incision-suture-time. Qualitative investigations in the medical workplace revealed to us, however, that other parts of the perioperative process are also relevant, in particular, induction and patient positioning. Including these steps in the analysis yields a more comprehensive picture of the perioperative process, and thus allowed for more precise duration estimations of the overall process. We conclude that a holistic view of medical processes is indispensable to effective overall improvements.

Including medical expertise in the analysis process proved vital in another aspect, too: the knowledge gathered from hospital staff considerably improved normalization of prose comments on medical records and operation plans, leading to better clustering and in turn much better predictive models. The hybrid approach of combining data science with expert input not only enabled more effective grouping but also led to a meaningful breakdown of the process into finer steps that are operationally relevant.

In creating predictive models for operation duration, we identified a number of relevant factors such as type of positioning and induction, type of intervention, and patient characteristics such as age, sex, and weight. Regression analyses and hypothesis tests indicate, however, that patient characteristics have much less impact than other factors. Thus, practical predictions of operation durations do not require personal data, which potentially affect patient privacy rights. Good predictive performance can be achieved with just a few well-prepared features, which means, that efficient models can be built in data-scarce environments, too.

Nevertheless, we suggest that there are additional factors that likely influence procedure duration and should be included in future analyses. These include e.\ g., comorbidities, the general health of the patient, and medications currently being taken. Most clinical experts also believe that the surgeon can significantly influence the duration of the procedure.

We believe that our findings are not restricted to the particular hospital we have studied. While other hospitals may have other profiles in terms of the procedures they perform, there are only two likely factors to explain major differences in average operation durations per procedure across hospitals: differences in execution or differences in individual capabilities. Both of these are interesting in their own right. When different hospitals do things differently and there are differences in medical outcome or cost, it is well worth studying such differences to spread best practices more quickly. When individual practitioners perform better than others, this should prompt us to improve staff training, or allocation. Either way, analytical perspective based on medical practices holds significant improvements both in terms of economics, and medical outcomes.

\section{Threats to Validity}
\label{threats}

Given the nature of our study, several of the widely known threats to validity simply do not apply. Take \emph{internal validity} as an example. We have developed and analyzed a causal model for impact factors. While this model is plausible and straight-forward, and we believe in it, the model is not part of our conclusions. We only use it to create a predictive model, the effectiveness of which is undeniable. So, even if our model were wrong or incomplete, would this not affect our conclusions.

Similarly, \emph{construct validity} is not a relevant category for our study: we measure only time durations such as incision-suture time or induction time that represent Key Performance Indicators well established in medical practice and research literature.

\subsection{Conclusion Validity}

From a data science perspective, 16,576 data records may appear to be a small data set, such that conclusions are in danger of introducing potential bias. From a medical perspective, however, this is a very large data set documenting the work of more than 5,000 medical professionals for more than a year. Recall that the conclusions we draw from our data analysis are mostly qualitative, indicating process improvement potential rather than changes to medical interventions. Thus, we believe our conclusions are well covered by the amount of data we have based them on.

Another potential weakness of data based studies is data quality. %
It appears, that some teams and clinics have a relaxed attitude towards documenting medical procedures. This change results in some events relevant to our analysis being missing from a significant number of workflows. For this reason, we treat each sub-process separately rather than modeling the perioperative process as a single continuous flow. %
Fig.~\ref{data_overview} illustrates these data gaps. Specifically, 69.98\% of all workflows include both incision and suture events needed for modeling the procedure sub-process. 46.50\% contain both the anesthesia\_start  and the anesthesia\_complete timestamps and the associated attributes needed for modeling the induction step. Only 6.97\% of workflows include events and information necessary for estimating the duration of the preparation phase.

We excluded 812 outliers that, to the eyes of medical professionals we consulted, were clearly flukes, such as when the induction timestamp is after the operation, or when operation timestamps indicated operation durations of several days. While the latter is not impossible, it is exceedingly rare and of no import to our conclusions. Again, we believe that our conclusions are not affected by poor data quality.

\subsection{Ecologic Validity and Generalizability}

Our study was conducted using real (historical) data, guaranteeing a high level of ecological validity. However, the site of our study is a university hospital that acts as a medical hub in a major metropolitan region. Therefore, the LMU University Hospital is faced with a very large diversity of medical cases as well as emergencies. This is a marked difference to many smaller hospitals, that care for much fewer complex and unusual medical situations. Thus, in smaller hospitals, creating a reliable operation schedule is typically an easier task. Conversely, however, smaller hospitals also have fewer staff and operating rooms, so there is a greater need for precise planning. In summary, we cannot know at this point, whether smaller hospitals will benefit from our approach in the same way and to the same degree that a major medical hub like LMU does. Observe, however, that our factor model (cf.\ Fig.~\ref{factor_model}) does document these potential influences, showing arrows from the context factors to the patient and treatment factor groups, respectively.

While average durations for standard procedures may vary between hospitals, each individual hospital can easily collect the data required to make predictive models that exceed manual estimates in quality. Even small hospitals can aggregate enough data to achieve significant plan stability improvements.

\subsection{Ethical Considerations}

Our study has not involved patients or their treatment, so our study did not require formal clearance from an Ethical Review Board or similar. We worked exclusively with anonymized, historical data that did not allow us to identify individuals, so our study does not affect privacy rights of patients either. Consequently, no consent of patients was required beyond the consent implied in undergoing medical treatment, as documented by the hospital. The data did also not contain personal information about medical staff, so no labor protection regulations were affected. 

\section{Contribution and Conclusion}
\label{conclusion}

High plan stability is a major goal of any hospital, as it affects economic as well as medical outcomes. Medical contingencies and ad hoc patient logistics aside, the only controllable factor for optimizing plan stability is the precision of effort estimations of procedures that are used to create the operation schedule. 

We have studied medical data from a large and renowned university hospital with a high degree of digitalization, covering the work of more than 5,000 medical professionals for over a year. Yet, the estimates of operation duration are created manually, exhibiting artifacts and bearing little correlation to actual operation durations. Replacing manual estimates even by trivial estimation models like the arithmetic mean of historic durations of comparable procedures offers significant improvements of effort estimations. Our findings transfer to other settings, though the impact might differ depending on the profile of procedures executed at a given hospital.

By combining expert knowledge, data-driven hypothesis testing, statistical validation, process modeling, and clustering, we gained a deeper understanding of the perioperative workflow and significantly improved prediction accuracy and planning reliability.

From a data science perspective, we are working with a small data set. By integrating qualitative research methods to include human expertise from the medical professionals involved in the planning process, we were able to leverage this small data set to achieve major improvements. So, a small dataset \emph{may} go a long way, sometimes.

%
%
\bibliographystyle{splncs04}
\bibliography{references,LiteraturUTF8}

\end{document}